\documentclass[10pt,twocolumn,letterpaper]{article}

\usepackage[T1]{fontenc}
\usepackage[pagenumbers]{cvpr}
\usepackage[dvipsnames]{xcolor}
\usepackage{amsmath,amssymb,booktabs,graphicx,microtype,xspace,array,tabularx}
\newcolumntype{Y}{>{\raggedright\arraybackslash}X}

\newcommand{\method}{\textnormal{\textsc{SubDGuide}}\xspace}
\newcommand{\target}{\mathcal{M}_{T}}
\newcommand{\cage}{\mathcal{C}}

\definecolor{cvprblue}{rgb}{0.21,0.49,0.74}
\usepackage[breaklinks,colorlinks,citecolor=cvprblue,linkcolor=cvprblue,urlcolor=cvprblue]{hyperref}

\title{SubDGuide: A Modeler-Inspired Agentic Workflow for Mesh-to-SubD Reconstruction}

\author{Mengnan Jiang$^{1,2}$ \quad Christian Franke$^{1}$ \quad Michele Franco Adesso$^{1}$\\
Antonio Haas$^{1}$ \quad Grace Li Zhang$^{2}$\\[4pt]
$^{1}$Mercedes-Benz AG \quad $^{2}$Technical University of Darmstadt\\
{\scriptsize\texttt{mengnan.jiang@mercedes-benz.com}\quad\texttt{christian.franke@mercedes-benz.com}}\\
{\scriptsize\texttt{michele\_franco.adesso@mercedes-benz.com}\quad\texttt{antonio.haas@mercedes-benz.com}}\\
{\scriptsize\texttt{grace.zhang@tu-darmstadt.de}}}

\begin{document}
\maketitle

\begin{abstract}
Subdivision surfaces represent free-form geometry through a sparse control cage, but recovering that cage from a dense mesh is not merely a fitting problem: the system must infer where control is needed, which curves encode design features, and when an initial result should be revised. We present \method, a modeler-inspired agentic workflow for this task. Stage A reads aligned multiview geometric evidence and produces a compact plan for cage resolution, feature mapping, and broad-form fitting. Stage B inspects the resulting surface, requests targeted diagnostics, and chooses repair, rollback, or stopping actions. Geometry tools execute and verify every change; the planner never generates vertices or connectivity.

On a fixed evaluation cohort, \method improves all five reported metrics over automatic remeshing: median Chamfer-L1 decreases from \(0.641\%\) to \(0.443\%\) of the target bounding-box diagonal, and F-score at a \(1\%\) tolerance increases from \(82.00\%\) to \(92.78\%\). Stateful feedback improves four of five reported median metrics over one-shot planning, while verification retains the earlier checkpoint when a proposal is unhelpful. The same interface supports six multimodal planners, showing how semantic judgment can guide a practical, inspectable mesh-to-SubD workflow.
\end{abstract}

\section{Introduction}
\label{sec:intro}

A dense triangle mesh records where a surface lies, but not how it was designed. A subdivision model adds a sparse control cage, explicit boundaries and creases, and an edge flow that determines how the shape behaves after refinement. Recovering this hidden structure is underdetermined: many cages can approximate the same samples, yet only some preserve the features that define the object.

\begin{figure}[t]
    \centering
    \includegraphics[width=0.9\linewidth]{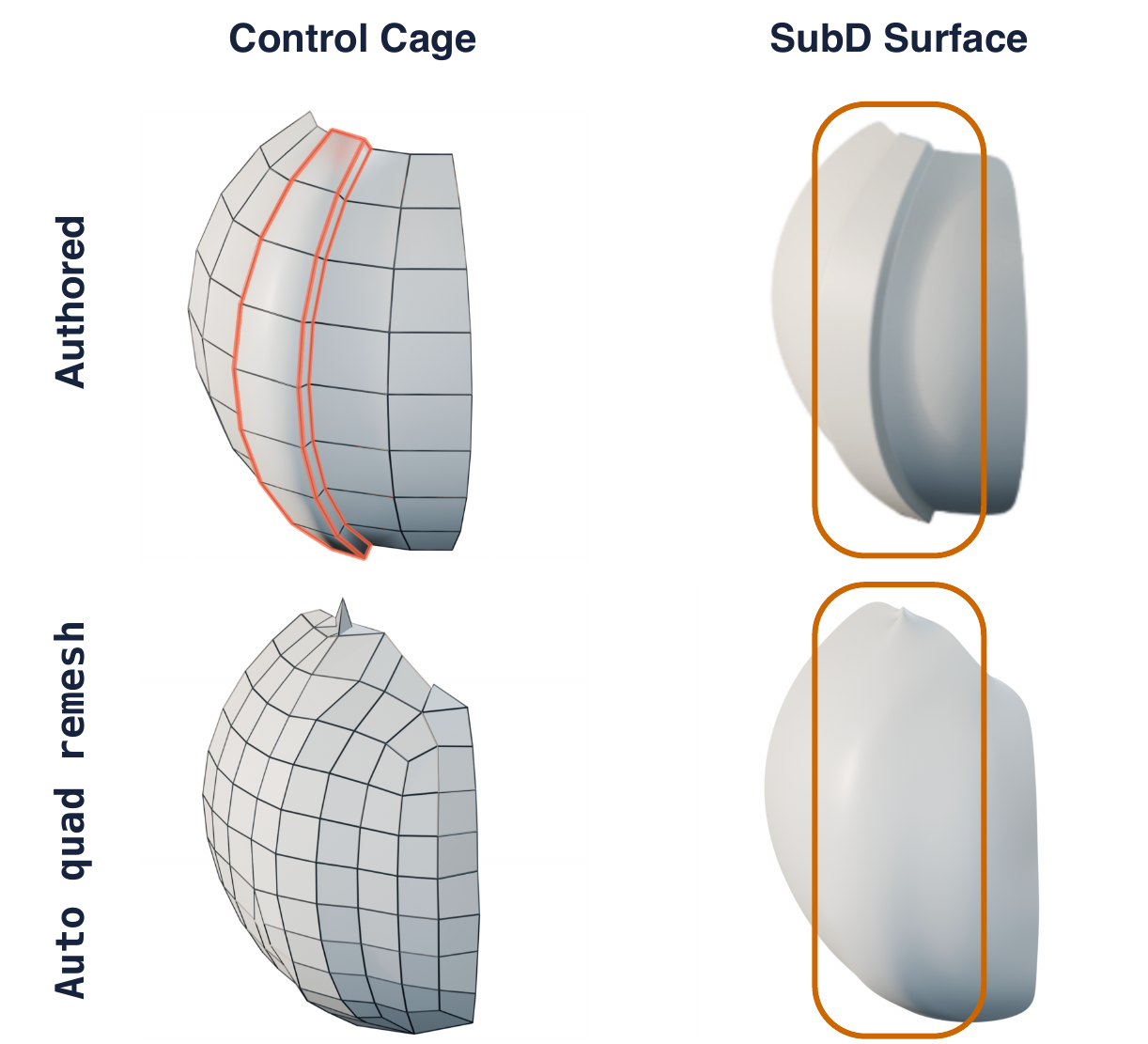}
    \caption{Similar geometry can hide different control structure. The authored cage supports the raised ridge, whereas automatic quad remeshing recovers the broad form but smooths the feature away after subdivision.}
    \label{fig:intro-ambiguity}
\end{figure}

Figure~\ref{fig:intro-ambiguity} shows the central ambiguity. The automatic result recovers the silhouette and broad volume, yet smooths away the narrow ridge because its control flow does not support that feature. The same problem appears at valleys, open boundaries, returns, and transitions between smooth and sharp regions. Geometric proximity alone therefore does not tell us whether a reconstructed cage expresses the target in the right way.

\begin{figure*}[t]
    \centering
    \includegraphics[width=0.95\textwidth]{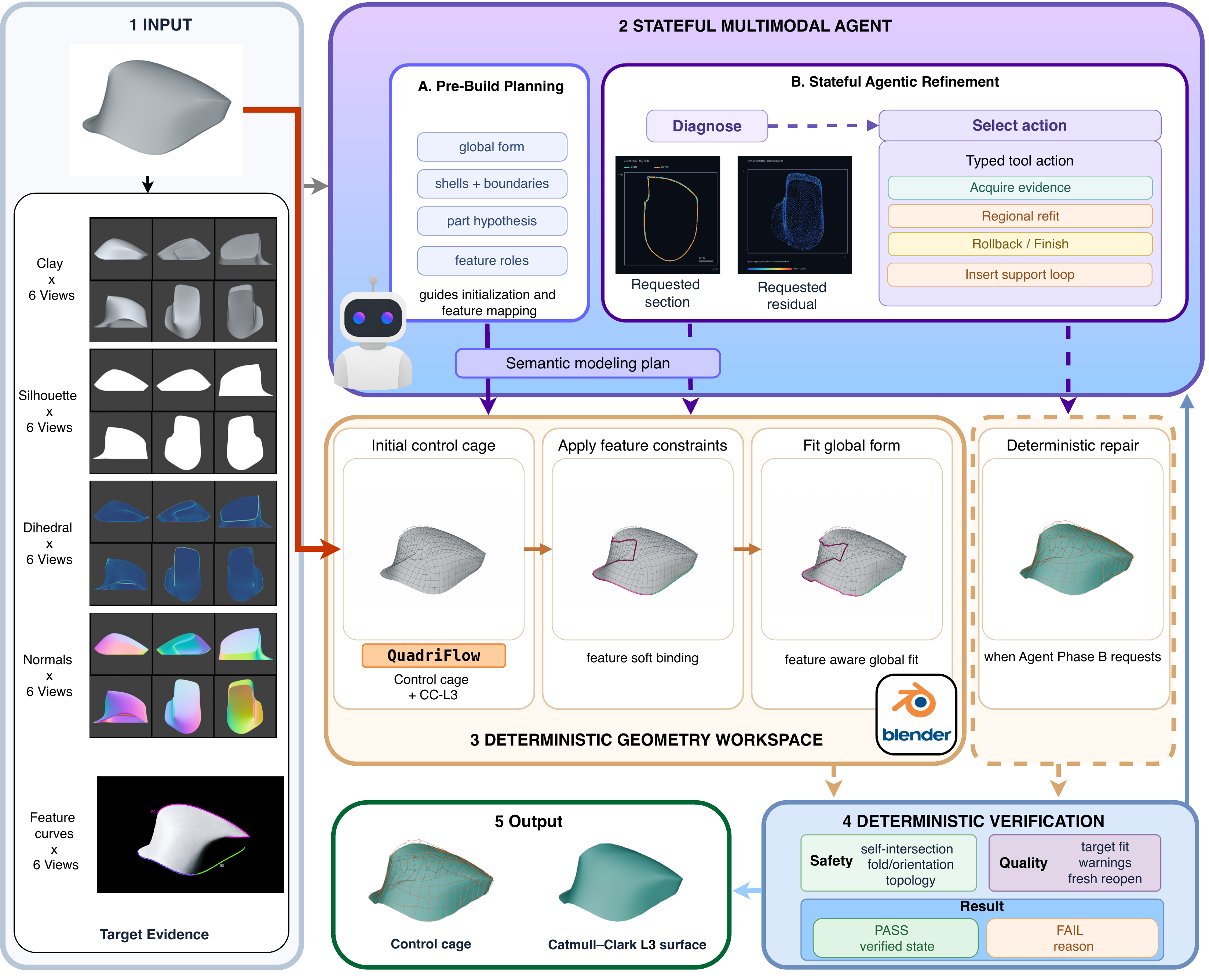}
    \caption{Overview of \method. Multiview evidence drives bounded pre-build planning; deterministic tools construct, fit, and verify the cage. A feedback controller may request additional evidence, apply a supported repair, revisit a checkpoint, or return the best verified result.}
    \label{fig:overview}
\end{figure*}

Professional retopology handles this ambiguity as a sequence of decisions. A modeler studies the object from several views, separates broad form from persistent features, chooses a suitable control density, blocks out the cage, and only then refines local structure. If the resulting surface exposes a wrong profile or poor edge flow, the modeler inspects a more diagnostic view and revisits the responsible step. This ordering matters: fixed-topology fitting can improve a suitable block-out, but it cannot invent a missing loop or rescue a fundamentally unsuitable initialization.

Numerical methods already provide strong quadrangulation and subdivision-fitting components~\cite{kanai2001meshtoss,litke2001fitting,ma2004direct,lavoue2007framework,lavoue2009semisharp,huang2018quadriflow,xie2022differentiable}. Multimodal language models offer a complementary ability: they can compare several views, interpret the role of a curve, and choose among structured modeling alternatives. Asking them to emit vertices or connectivity directly, however, entangles visual judgment with numerical construction and leaves little room for verification. Our premise is simpler: the planner decides \emph{what should be attempted}; geometry tools decide \emph{how it is constructed and whether it works}.

We therefore introduce \method, summarized in Fig.~\ref{fig:overview}. \emph{Target Evidence} combines aligned shape and feature views. In \emph{Stage A: Pre-Build Planning}, the multimodal planner chooses a scale-aware initialization and assigns a role to each candidate feature. The \emph{Deterministic Geometry Workspace} constructs the cage, maps selected curves to real cage-edge paths, and performs feature-aware global fitting. In \emph{Stage B: Stateful Agentic Refinement}, the planner reviews the actual CC-L3 result, requests a residual or section when needed, and invokes a supported repair or returns to an earlier state. Finally, \emph{Deterministic Verification} independently rebuilds and checks every candidate before the best verified result is returned.

We evaluate five controlled system configurations on a fixed 20-shape cohort. The configurations vary along two axes: how the pre-build plan is produced and whether post-build feedback is available. Relative to automatic remeshing, the complete workflow improves every reported reconstruction metric. Its feedback loop also produces verified geometric changes rather than comments alone: both directional distances improve in six cases, while the earlier result is preserved in the other 14. Replacing the planner with five alternative multimodal models produces closely clustered geometry but different search behavior, supporting the intended separation between a replaceable semantic planner and a fixed geometric workspace.

Our contributions are:
\begin{itemize}
    \item a two-stage, modeler-inspired Mesh-to-SubD workflow that connects multiview interpretation, cage construction, feature-aware fitting, and feedback-driven repair;
    \item an executable interface through which semantic decisions change initialization, real cage-edge paths, and supported repair operations without asking the planner to generate geometry; and
    \item a 20-case evaluation of five controlled configurations, including direct measurements of when feedback changes geometry and how six planners use the same workflow.
\end{itemize}

\section{Related Work}
\label{sec:related}

\paragraph{Triangle surfaces as upstream geometry.}
Classical reconstruction converts unorganized samples into piecewise-smooth or watertight triangle surfaces~\cite{hoppe1992surface,hoppe1994piecewise,kazhdan2006poisson,kazhdan2013screened}. Learned methods predict meshes or implicit fields from images and point sets~\cite{wang2018pixel2mesh,dai2019scan2mesh,park2019deepsdf,mescheder2019occupancy}, while generative systems increasingly produce explicit 3D geometry~\cite{nash2020polygen,shen2021dmtet,gao2022get3d,poole2023dreamfusion,lin2023magic3d,jun2023shapee,siddiqui2024meshgpt,chen2024meshanything,chen2025meshanythingv2,fang2025meshllm,xiang2025trellis}. These methods improve geometric or visual fidelity, but their output does not by itself reveal the sparse control structure from which a feature-aware subdivision surface could be edited or reproduced. We treat scanned, modeled, or generated triangle meshes as possible upstream inputs to inverse reconstruction.

\paragraph{Structured CAD and free-form reverse engineering.}
Reverse engineering has long sought compact design structure rather than only dense samples~\cite{varady1997reverse,krishnamurthy1996fitting,eck1996automatic,koch2019abc}. Learned systems recover primitives, sketches, extrusion histories, zones, and boundary representations~\cite{wu2021deepcad,willis2021brepnet,willis2021fusion360,xu2021zonegraphs,uy2022point2cyl,xu2022skexgen,li2023secadnet,khan2024cadsignet,chen2025cadcrafter,rukhovich2025cadrecode,li2025caddreamer}. Free-form methods instead segment and fit analytic or spline patches, assemble B-Rep complexes, or fit NURBS surfaces~\cite{sharma2020parsenet,guo2022complexgen,liu2024point2cad,fuentesperez2026nurbsfit}. Our target is complementary: a sparse Catmull--Clark control cage whose broad form, boundaries, and salient free-form features agree with the input mesh.

\paragraph{Subdivision surfaces and quadrilateral control structure.}
Catmull--Clark and related rules provide smooth arbitrary-topology surfaces with local control~\cite{catmull1978recursively,doo1978behaviour,halstead1993efficient,derose1998subdivision,stam1998exact,biermann2000piecewise,opensubdiv}. Prior inverse-subdivision work fits control meshes to points or dense polygonal surfaces~\cite{kanai2001meshtoss,litke2001fitting,cheng2004sdm,ma2004direct,marinov2005optimization,lavoue2007framework,lavoue2009semisharp,xie2022differentiable}. Quadrangulation supplies field-aligned, feature-aware, or learned initial layouts~\cite{bommes2009miq,bommes2012quadmeshing,jakob2015instant,huang2018quadriflow,pietroni2021featureline,liu2025neuframeq,kong2026squadgen}; industrial tools likewise expose SubD alignment, creasing, retopology, and Mesh-to-SubD conversion~\cite{autodesk2024aliassubd,autodesk2026aliasmesh}. These components are essential, but they do not decide how several ambiguous views should influence control density, feature semantics, or the decision to revise an already built cage.

\paragraph{Agentic 3D creation and editing.}
Language- and vision-conditioned systems plan scenes, emit Blender programs, arrange assets, and revise graphics through tool feedback~\cite{sun2023threedgpt,hu2024scenecraft,huang2024blenderalchemy,yang2024holodeck,du2024blenderllm,gu2025blendergym,lu2025ll3m,wang2026adsl}. Their action spaces demonstrate that semantic planning can guide executable 3D operations. They do not, however, directly address inverse recovery of an explicit Catmull--Clark cage from a dense input mesh, where a visually plausible suggestion may still lead to poor edge flow or an invalid subdivided surface.

\paragraph{Language and agent interfaces for CAD.}
Language-conditioned CAD spans text-to-CAD, vision-to-code, semantic annotation, parametric sequence generation, and tool-assisted editing~\cite{badagabettu2024query2cad,alrashedy2025cadcodeverify,yuan2024cadtalk,li2025cadllama,chen2025cadcrafter,rukhovich2025cadrecode,mallis2025cadassistant}. Recent systems add iterative checking, relation extraction, multi-agent validation, or program search~\cite{yuan2026cadrefiner,kabisov2026cadreasoner,barkley2026cadsmith,ataei2026zerotocad,wang2026adsl,adler2026aadvark,zhang2026cadloop,ocker2025ideatocad}. Most operate on parametric CAD, solids, assemblies, or scripting representations. \method occupies a different intersection: the planner receives visual evidence from a free-form mesh, but can affect geometry only through a small set of deterministic cage-construction, mapping, fitting, and repair tools.

\section{Method}
\label{sec:method}

\begin{figure*}[t]
    \centering
    \includegraphics[width=0.95\textwidth]{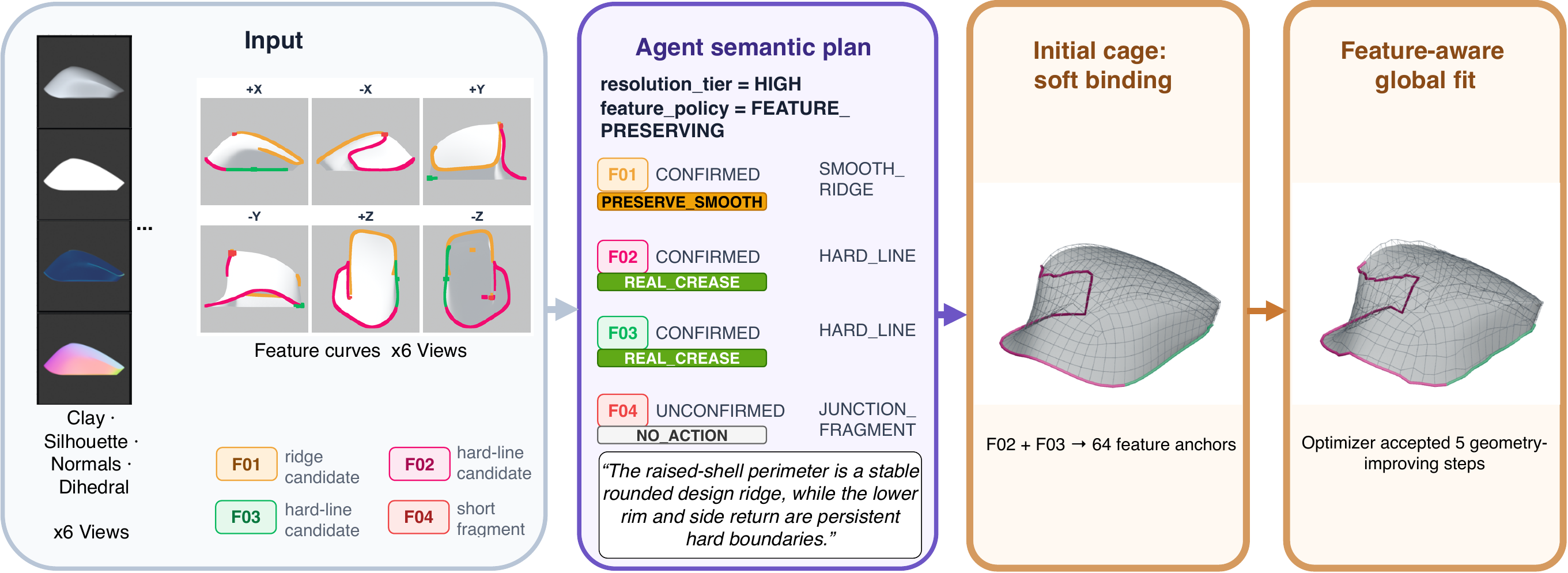}
    \caption{Stage~A on the Mouse target. The labels shown are the normalized planner output from the recorded run; the mapper turns these feature decisions into soft anchors and real crease-edge chains. This independent mechanism example is not part of the 20-case quantitative cohort.}
    \label{fig:stage-a-mouse}
\end{figure*}

\begin{figure*}[t]
    \centering
    \includegraphics[width=0.88\textwidth]{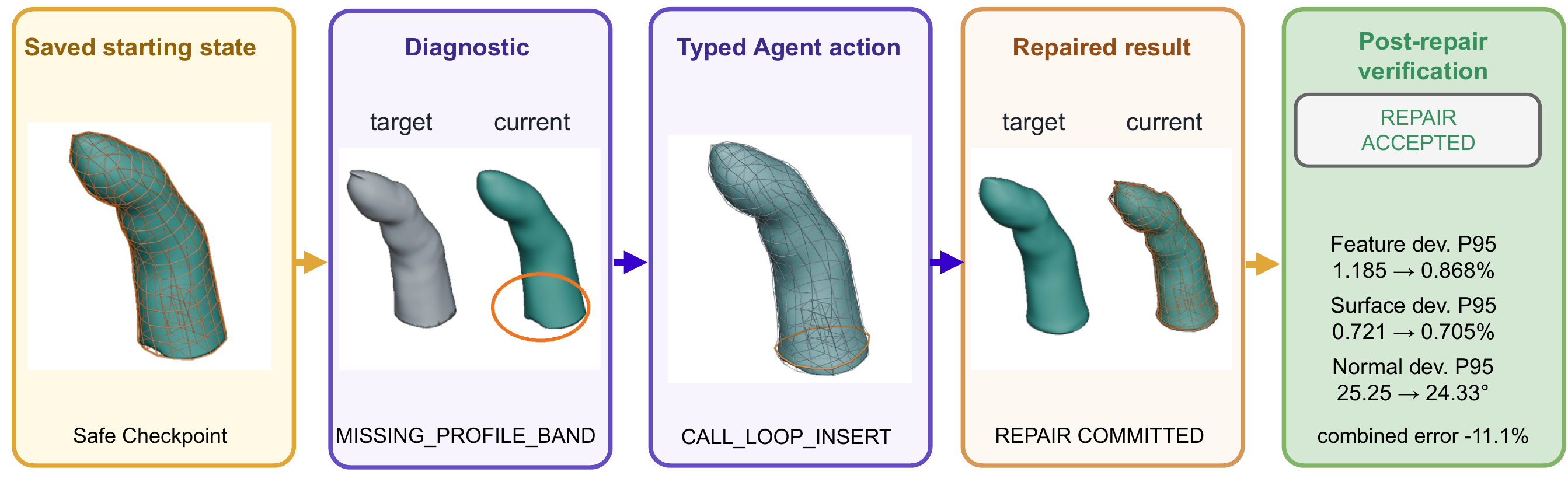}
    \caption{Stage~B on a curved shell. The displayed diagnosis and loop-insertion action come from the recorded controller step; the geometry tools construct, fit, and verify the retained result.}
    \label{fig:phaseb-trace}
\end{figure*}

\subsection{Problem and Workflow}

Given a triangle mesh $\target$, our goal is a compact quadrilateral cage $\cage$ whose subdivided surface follows the target while preserving its components, open boundaries, apertures, and salient curves. The problem is not to recover an unknowable unique source cage. It is to construct a valid control representation whose broad form and explicitly modeled features agree with the observed object.

Figure~\ref{fig:overview} organizes the method around the same sequence used throughout the paper: \emph{Target Evidence}, \emph{Stage A: Pre-Build Planning}, the \emph{Deterministic Geometry Workspace}, \emph{Stage B: Stateful Agentic Refinement}, and \emph{Deterministic Verification}. The planner interprets evidence and selects typed operations. Deterministic tools own every control-vertex position, edge connection, fitting update, and topology edit.

We normalize the target by its axis-aligned bounding-box diagonal $D$. The output is a sparse quadrilateral cage with boundary and crease attributes. Unless stated otherwise, we evaluate the surface after three Catmull--Clark steps~\cite{catmull1978recursively}, denoted \emph{CC-L3}; level four provides an additional consistency check. Checkpoint selection uses these reconstructed surfaces rather than an optimizer's internal surrogate.

\subsection{Stage A: Pre-Build Planning}

Six canonical cameras observe the target. Each camera produces a clay render, silhouette, world-space normal map, and dihedral-angle cue. A curve detector groups persistent boundary and high-dihedral edges into candidates and projects the same color-coded curves into all six cameras. The resulting evidence is spatially aligned: a curve in the feature overlay can be checked against silhouette, normal, and dihedral evidence from the same viewpoint.

The planner also receives compact, normalized descriptors: component and boundary summaries plus, for each candidate, length, location, endpoints, closure, branching, and per-view visibility. These descriptors provide the shell, boundary, and part evidence summarized in Fig.~\ref{fig:overview}; they do not expose raw mesh arrays or cage connectivity. Stage~A returns a low, medium, or high resolution tier, an initialization policy, and exactly one semantic decision for each candidate. A candidate may become a crease, remain a smooth feature, preserve an open boundary, or be ignored.

These choices are small but geometrically consequential. Resolution and feature policy configure the initial quadrangulation; curve decisions determine which structures constrain fitting and which mapped paths later receive crease attributes. Figure~\ref{fig:stage-a-mouse} shows a real Stage-A trace: the planner distinguishes a smooth ridge from two hard lines, while the following geometry stages turn those labels into anchors and crease-edge chains.

\subsection{Deterministic Geometry Workspace}

A QuadriFlow wrapper constructs the initial all-quad cage using the Stage-A resolution and feature policy. The base budget scales with observed curve complexity; the selected tier changes that budget instead of choosing an object-specific template. Input components, genus, and open boundaries are carried into construction and checked before fitting. If the selected tier fails to produce a valid start, one predefined lower tier is available. This stage is load-bearing because later fitting can move existing control vertices but cannot invent missing edge flow.

For every selected ridge or hard line, the feature mapper searches the actual cage graph for an edge path that follows the target curve. The path cost balances spatial distance, tangent agreement, forward progress, and penalties for backtracking or jumping across curve segments. An unmappable candidate is omitted without invalidating mappings that already succeed.

Mapping first acts as a \emph{soft constraint} on the initial cage. Feature-aware global fitting minimizes
\begin{equation}
E_{\mathrm{fit}}(\cage)=
w_s E_{\mathrm{surf}}+w_v E_{\mathrm{sil}}+
w_f E_{\mathrm{feat}}+w_r E_{\mathrm{reg}},
\label{eq:fit-objective}
\end{equation}
where the terms align the subdivided surface, six-view silhouettes, and mapped features while regularizing displacement and cage distortion. All weights and step limits are fixed across the evaluation. This ordering leaves enough freedom to establish the global form before any edge is hardened.

Because fitting moves the cage, the target curves are mapped again on the fitted graph. Curves labeled as hard lines are written to real cage edges as crease attributes; smooth features keep their alignment role without being hardened. A local refinement then improves positional and tangent agreement around the mapped paths, and Blender evaluates the resulting crease-aware CC-L3 surface.

The planner never draws the colored cage chains in Fig.~\ref{fig:stage-a-mouse}. It identifies target features and their intended treatment; the mapper finds executable edge paths, and the fitter uses those paths as real constraints. This makes the effect of Stage~A traceable from target evidence to the final cage.

\subsection{Stage B: Stateful Agentic Refinement}

The first reconstructed surface becomes the starting state for Stage~B. At each review, the controller sees the target, current CC-L3 surface, cage overlay, residuals, the previous tool result, earlier verified states, and its remaining budget. It may first acquire evidence through updated multiview renders, a localized residual close-up, or a triangle--plane section.

The geometry actions match the verbs in Fig.~\ref{fig:overview}. \emph{Regional refit} adjusts a diagnosed area without changing connectivity. \emph{Repair feature/topology} currently offers one local topology operation: when the detector finds an admissible finite-width site, the tool inserts a support loop and refits the affected region. The controller may also roll back to the initial or globally fitted state, finish with the best verified result, or stop without returning geometry. It cannot synthesize arbitrary topology, rewrite the mapper, or turn a failed verification into a pass.

Each proposed repair is evaluated before it can replace the current best state. With verification indicator $V$ and comparison score $S$, the update is
\begin{equation}
\cage_{\mathrm{best}}\leftarrow\cage'
\quad\text{iff}\quad
V(\cage')=1\;\land\;S(\cage')<S(\cage_{\mathrm{best}}).
\label{eq:checkpoint-rule}
\end{equation}
Otherwise the proposal remains visible in the trace but the earlier result is retained. Feedback can therefore discover a better executable state without forcing an unhelpful edit into the output.

Figure~\ref{fig:phaseb-trace} shows this loop in an executed case. The controller diagnoses a missing profile band and selects loop insertion. The tool constructs and fits the proposal, and verification decides whether the repaired state is retained.

\subsection{Deterministic Verification and Output}

The \emph{Deterministic Verification} module in Fig.~\ref{fig:overview} independently reconstructs every initial, fitted, or repaired cage in Blender, including its boundary and crease attributes. It checks cage degeneracy, CC-L3 folds and self-intersections, orientation and Jacobian indicators, component and aperture preservation, bidirectional distance, normal agreement, and confirmed-feature error.

At termination, all retained states are rebuilt once more and the best verification-passing result is returned. If none passes, the workflow returns no geometry. The output contains the control cage, boundary and crease attributes, the evaluated CC-L3 surface, and the verification report. Because each transition records its observation, typed action, tool result, and resulting state, a semantic decision can be followed to a concrete geometric effect.

\section{Experiments}
\label{sec:experiments}

The experiments follow the method's three claims. First, we test whether Stage~A planning and the geometry workspace improve reconstruction over automatic remeshing. Second, we isolate what Stage~B feedback changes after the first build. Third, we replace the multimodal planner while holding the rest of the workflow fixed.

\subsection{Setup}

We use a fixed cohort of 20 single-component meshes: sixteen from Thingi10K~\cite{zhou2016thingi10k} and four pre-existing anonymous reference shapes from our controlled development suite. The set emphasizes moderately complex free-form shells, elongated parts, shallow enclosures, open boundaries, and limited apertures. All five configurations produce comparable outputs on this cohort. Every planner receives the same anonymized Target Evidence used in Stage~A: rendered views and normalized descriptors, but no raw mesh arrays or source SubD cage. The same target normalization, initial budget rule, fitting limits, and verification thresholds are used throughout; the exact cohort is listed in the supplement.

We compare five controlled system configurations under the same target, cage-budget policy, fitter, and CC-L3 evaluator. They vary along two axes: the source of the pre-build plan and the availability of post-build feedback. The Rule-based Controller is the strongest fully deterministic reference, while the comparison between One-shot AI Planning and Ours isolates Stage~B feedback under the same Stage~A planner:

\begin{itemize}
    \item \textbf{Automatic Remeshing} constructs and evaluates the complexity-scaled QuadriFlow cage;
    \item \textbf{Deterministic Fitting} adds the fixed global and feature-aware fitting sequence;
    \item \textbf{Rule-based Controller} chooses from the same evidence and repair tools using fixed rules;
    \item \textbf{One-shot AI Planning} uses Stage~A multimodal planning, then runs the geometry workspace once; and
    \item \textbf{Ours} adds Stage~B evidence acquisition, repair, rollback, and verified output selection.
\end{itemize}

One-shot AI Planning and Ours use GPT-5.6 Sol at high reasoning effort. Each configuration is run once per case. Stage~B allows at most 15 reviews and ten geometry attempts within 600 seconds. For the cross-model study, we change only the planner: GPT-5.6 Sol, Terra, or Luna; Claude Opus~5; Gemini 3.7 Flash; or Kimi K3. The prompt, evidence, action vocabulary, geometry tools, verification, and budgets remain fixed. All multimodal models are accessed through standard hosted APIs; no model is trained or fine-tuned for this work. Geometry processing, Blender reconstruction, and evaluation run locally on a MacBook Pro with an Apple M3 Max and 64\,GB unified memory.

\begin{figure*}[t]
    \centering
    \includegraphics[width=0.98\textwidth]{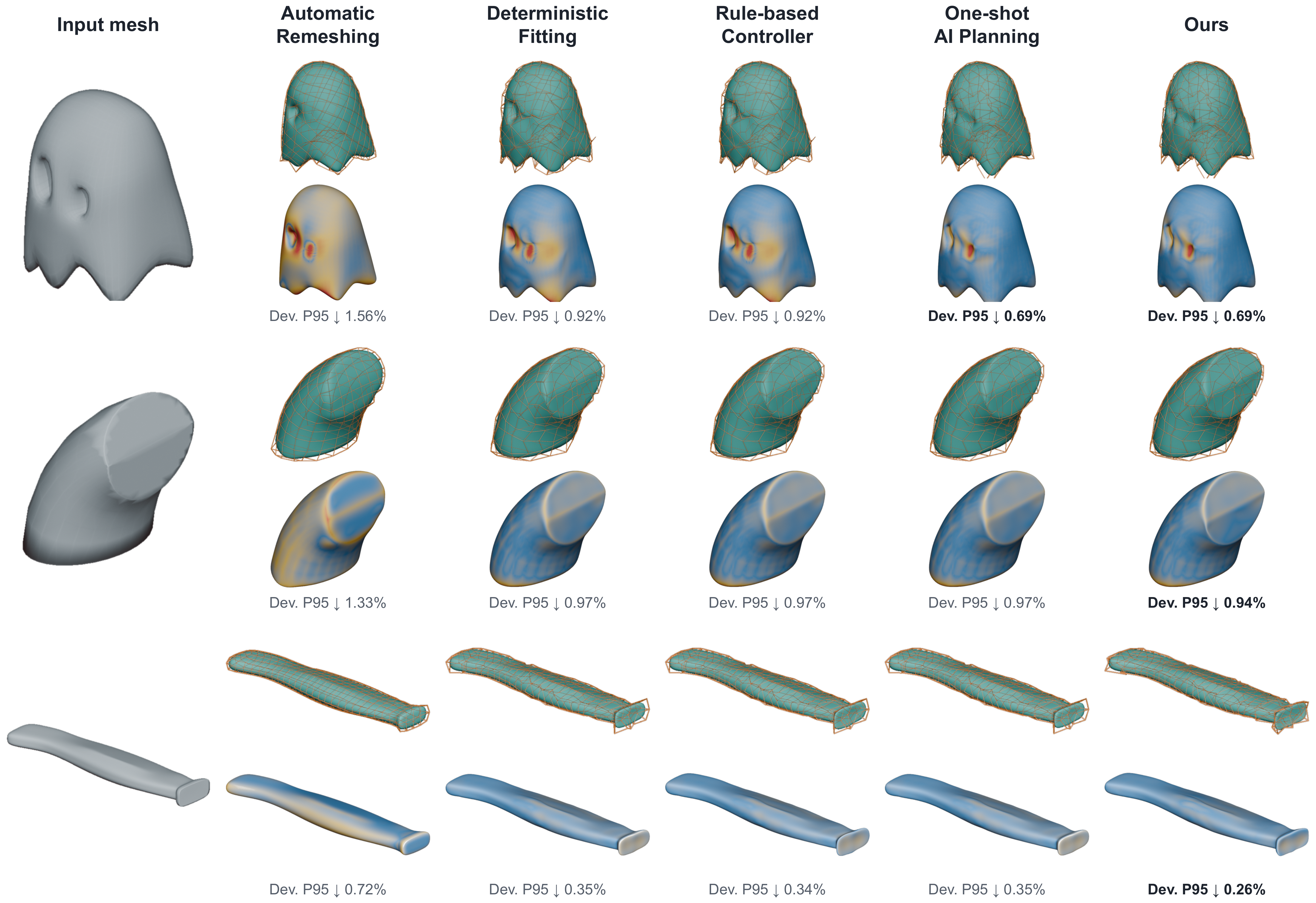}
    \caption{Representative reconstructions from the 20-case evaluation. Each panel shows the retained CC-L3 surface and control cage; color encodes output-to-target deviation on a shared scale from 0 to 2\% of $D$.}
    \label{fig:qualitative}
\end{figure*}

\subsection{Metrics}

We draw 20,000 area-weighted samples from each target and evaluated output using a fixed sampler. Every distance is reported as a percentage of the target's axis-aligned bounding-box diagonal $D$. A P95 value is the 95th percentile: 95\% of the measured samples have an error no larger than that value, so it reflects difficult local regions without being dominated by a few isolated samples. \textbf{Dev. P95} applies this statistic from output to target and exposes protruding or displaced reconstruction. \textbf{Chamfer-L1} averages nearest-surface distance in both directions and summarizes overall closeness. \textbf{F-score} is evaluated at a tolerance equal to 1\% of $D$ and is high only when the output is both precise and covers the target within that tolerance. \textbf{Normal P95} measures the upper tail of surface-orientation error, while \textbf{six-view silhouette IoU} measures projected outline agreement. For the feedback audit we also retain target-to-output P95, which emphasizes missed target coverage. Table~\ref{tab:main-comparison} reports the case-wise median over the same 20 meshes.

\begin{table*}[t]
\centering
{\footnotesize
\renewcommand{\arraystretch}{1.08}
\begin{tabular*}{0.96\textwidth}{@{\extracolsep{\fill}}lccccc@{}}
\toprule
Configuration & \shortstack{Dev. P95\\(\% of $D$) $\downarrow$} & \shortstack{Normal P95\\($^\circ$) $\downarrow$} & \shortstack{Chamfer-L1\\(\% of $D$) $\downarrow$} & \shortstack{Surface F-score\\@1\% of $D$ (\%) $\uparrow$} & \shortstack{6-view Silhouette\\IoU (\%) $\uparrow$} \\
\midrule
Automatic Remeshing   & 1.242 & 36.92 & 0.641 & 82.00 & 87.73 \\
\midrule
Deterministic Fitting & 1.037 & 32.90 & 0.462 & 89.91 & 91.13 \\
Rule-based Controller & 0.996 & 32.94 & 0.462 & 91.27 & 91.26 \\
One-shot AI Planning  & 0.975 & 29.71 & \textbf{0.443} & 92.62 & 92.96 \\
\midrule
Ours                  & \textbf{0.970} & \textbf{29.34} & \textbf{0.443} & \textbf{92.78} & \textbf{93.13} \\
\bottomrule
\end{tabular*}
}
\caption{Geometry comparison on the fixed 20-case cohort. Automatic Remeshing is the external baseline; the middle rows are controlled variants of the shared geometry workflow. Values are case-wise medians. Figure~\ref{fig:qualitative} visualizes the same Dev. P95 measure.}
\label{tab:main-comparison}
\end{table*}

\subsection{Reconstruction Quality}

Table~\ref{tab:main-comparison} shows the progression. Deterministic Fitting produces the largest improvement over Automatic Remeshing, reducing median Dev. P95 from 1.242\% to 1.037\% of $D$ and Chamfer-L1 from 0.641\% to 0.462\% of $D$. This confirms that generic quadrangulation can provide useful connectivity, but its CC surface still requires target-aware geometric alignment. The Rule-based Controller gives a further reduction in the distance tail. Stage~A planning then improves normal, F-score, and silhouette metrics by adapting resolution, initialization policy, and feature treatment.

One-shot AI Planning is the single-pass reference configuration. It receives the complete Stage-A evidence and produces the same semantic plan used by Ours, after which the workspace builds and fits the cage once. What it lacks is Stage~B: it cannot inspect that result, request a more diagnostic view, repair it, or return to another verified state. Ours adds this feedback and obtains the lowest Dev. P95 (0.970\% of $D$) and normal P95 (29.34$^\circ$), the highest F-score (92.78\%) and silhouette IoU (93.13\%), and ties One-shot AI Planning on Chamfer-L1. Relative to Automatic Remeshing, Ours reduces median Dev. P95 by 21.9\% and Chamfer-L1 by 30.9\%, while increasing F-score from 82.00\% to 92.78\%. The smaller gap to the one-shot reference isolates the additional value of iterative feedback.

Figure~\ref{fig:qualitative} shows where the numerical progression comes from. Automatic Remeshing usually captures gross volume, but its evaluated surface visibly drifts around narrow lower boundaries, tapered ends, and transitions between rounded panels and sharper cuts. Deterministic Fitting removes much of the broad low-frequency mismatch. The feature-aware configurations further concentrate the remaining error into smaller regions and retain cleaner boundary and ridge behavior on the free-form shells. In the displayed cases, Ours is never worse than One-shot AI Planning and produces the clearest additional improvement on the elongated part and localized edge transitions. The visual difference between these last two columns is necessarily smaller than the difference from Automatic Remeshing, matching the quantitative ablation.

\subsection{What Feedback Changes}

\begin{figure}[t]
  \centering
  \includegraphics[width=0.98\columnwidth]{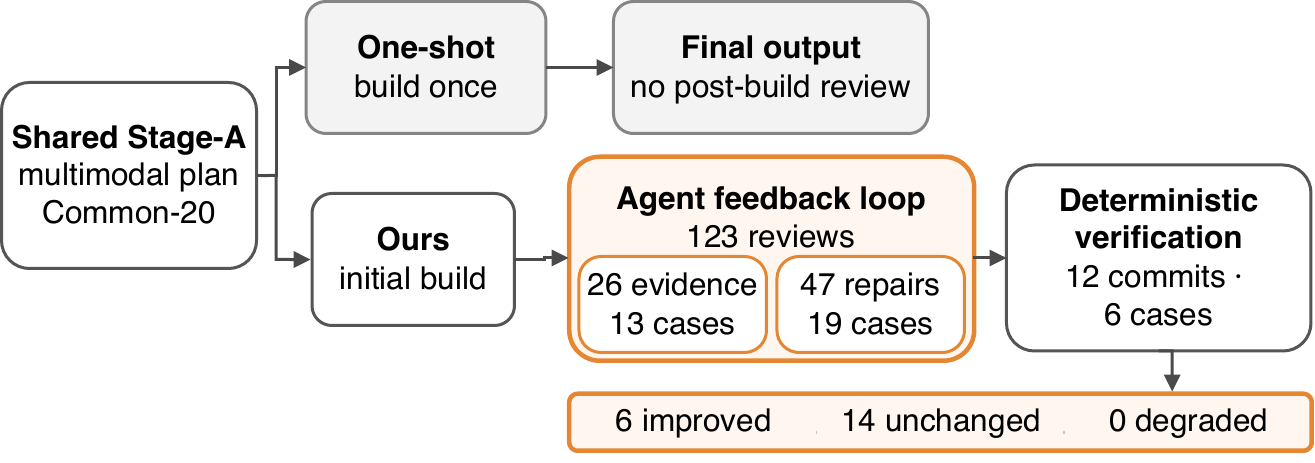}
  \caption{Effect of iterative feedback. Both configurations share the pre-build plan; Ours continues reviewing and commits only verified improvements. Six cases improve and 14 retain their previous geometry.}
  \label{fig:feedback-ablation}
\end{figure}

One-shot AI Planning and Ours share Target Evidence, Stage~A, the initial cage, feature mapper, fitter, and Deterministic Verification. The one-shot configuration ends after its first reconstructed result. Ours continues through Stage~B, where the model reviews the actual surface rather than its earlier prediction. It may request a residual close-up, section, or updated multiview render; invoke regional refitting or the supported loop insertion; or return to an earlier verified state. Figure~\ref{fig:feedback-ablation} summarizes these two paths.

Across the 20 cases, Stage~B performs 123 reviews, requests additional evidence 26 times in 13 cases, and evaluates 47 repair candidates in 19 cases. Deterministic Verification admits 12 geometry updates across six cases. In each of those six, both output-to-target and target-to-output P95 improve. The remaining 14 outputs are unchanged because Eq.~\ref{eq:checkpoint-rule} retains the earlier state when no supported repair is better. Zero degradation is therefore an intended consequence of the selection rule; the empirical result is that the loop finds a better executable state in six cases.

\subsection{Planner Choice and Efficiency}

Finally, we vary only the multimodal planner. Tables~\ref{tab:crossmodel-quality} and~\ref{tab:crossmodel-efficiency} separate the geometry retained by the shared workspace from the search behavior used to reach it. Geometry metrics use the retained output on all 20 cases for every planner, including the previous verified state when a provider reply fails. Search-efficiency statistics use only cases with at least one valid reply: 20 for GPT and Claude, and 15 for Gemini and Kimi.

\begin{table*}[t]
\centering
{\footnotesize
\renewcommand{\arraystretch}{1.08}
\begin{tabular*}{0.96\textwidth}{@{\extracolsep{\fill}}lccccc@{}}
\toprule
Planner & \shortstack{Dev. P95\\(\% of $D$) $\downarrow$} & \shortstack{Normal P95\\($^\circ$) $\downarrow$} & \shortstack{Chamfer-L1\\(\% of $D$) $\downarrow$} & \shortstack{Surface F-score\\@1\% of $D$ (\%) $\uparrow$} & \shortstack{6-view Silhouette\\IoU (\%) $\uparrow$} \\
\midrule
GPT-5.6 Sol                   & 0.9704 & 29.34 & 0.4433 & 92.78 & 93.13 \\
GPT-5.6 Terra                 & 0.9732 & 29.34 & 0.4433 & 92.74 & 93.13 \\
GPT-5.6 Luna                  & 0.9767 & 29.39 & 0.4433 & 92.63 & 93.02 \\
Claude Opus 5                 & 0.9685 & 29.58 & 0.4428 & 92.62 & 92.92 \\
Gemini 3.7 Flash             & 0.9685 & 29.71 & 0.4433 & 92.62 & 92.96 \\
Kimi K3                      & 0.9683 & 29.56 & 0.4433 & 92.62 & 92.92 \\
\bottomrule
\end{tabular*}
}
\caption{End-to-end geometry with shared evidence, tools, and verification. Values are case-wise medians over all 20 retained outputs. For Gemini and Kimi, five provider failures retain the preceding verified state rather than a planner-specific edit.}
\label{tab:crossmodel-quality}
\end{table*}

\begin{table*}[t]
\centering
{\footnotesize
\renewcommand{\arraystretch}{1.08}
\begin{tabular*}{0.88\textwidth}{@{\extracolsep{\fill}}lcccc@{}}
\toprule
Planner & \shortstack{Runs selecting a\\verified improvement} & \shortstack{Review\\rounds $\downarrow$} & \shortstack{Relative token use\\(Sol $=1.0$) $\downarrow$} & \shortstack{Time/case\\(min) $\downarrow$} \\
\midrule
GPT-5.6 Sol                   & 5 & 6.15 & 1.00 & 7.15 \\
GPT-5.6 Terra                 & 6 & 7.60 & 1.15 & 6.52 \\
GPT-5.6 Luna                  & 6 & 5.75 & 0.91 & 5.21 \\
Claude Opus 5                 & 9 & 6.45 & 1.14 & 6.98 \\
Gemini 3.7 Flash$^{\dagger}$ & 5 & 5.73 & 0.88 & 5.95 \\
Kimi K3$^{\dagger}$          & 5 & 7.13 & 1.04 & 6.72 \\
\bottomrule
\end{tabular*}
}
\caption{Cross-model search behavior over valid-reply cases. The first column counts runs ending with a committed repair or better-checkpoint selection; it is distinct from the paired geometric comparison in Fig.~\ref{fig:feedback-ablation}. Relative token use is normalized to GPT-5.6 Sol; $^{\dagger}$ denotes 15 rather than 20 valid-reply cases.}
\label{tab:crossmodel-efficiency}
\end{table*}

Table~\ref{tab:crossmodel-quality} shows that final geometry remains tightly clustered at the end-to-end system level: different models lead individual columns, but no planner dominates every metric. Kimi has the lowest Dev. P95, Claude Opus~5 the lowest Chamfer-L1, and GPT-5.6 Sol or Terra lead the normal, F-score, and silhouette columns. Planner identity affects which evidence and action are selected, while the shared workspace and verification keep the retained geometry inside the same operating regime. The five fallback cases for Gemini and Kimi prevent provider failure from being mistaken for a new geometric result.

Table~\ref{tab:crossmodel-efficiency} reveals larger behavioral differences. Claude Opus~5 selects a verified improvement in the most runs, GPT-5.6 Luna is fastest, and Gemini uses the fewest reported tokens per valid reply. These endpoint counts are not the same statistic as the six paired cases in Fig.~\ref{fig:feedback-ablation}, where both directional P95 distances improve over One-shot AI Planning; GPT-5.6 Sol selects five verified improvements but has six paired geometric improvements. More reviews or tokens do not automatically imply better geometry: they describe how a planner searches the available action space. Provider tokenization and valid-response counts differ, so the comparison demonstrates interchangeability and practical trade-offs rather than a universal model ranking.

\section{Limitations and Outlook}
\label{sec:limitations}

The initial cage remains the main geometric bottleneck. Stage~A can change its resolution and feature policy, and Stage~B can refine an existing region or insert one supported loop, but neither stage can generally recover a missing component or redesign arbitrary edge flow. Broader part decomposition, alternative cage constructors, and pole-aware local relayout are natural extensions of the same tool-mediated design.

The 20 meshes span several industrially relevant shell forms but remain a curated cohort drawn from public and controlled sources, and no source control cages are available. Our evaluation therefore measures agreement with the target and preservation of observable structure, not recovery of an artist's unique topology. Pole placement and feature-loop continuity are reported only as structural diagnostics; we do not measure interactive editing performance. Each configuration is evaluated once per case, so the results characterize the frozen runs rather than repeated-call variability. Finally, cross-model token counts use provider-specific accounting and are best read as an operational efficiency measure rather than a hardware-independent cost.

\section{Conclusion}

\method combines multiview planning, cage construction, stateful feedback, and verification. It improves all five metrics over automatic remeshing across six planner models, without direct geometry generation.

{\small
\bibliographystyle{ieeenat_fullname}
\bibliography{main}
}

\clearpage
\appendix
\maketitlesupplementary
\section{Evaluation Protocol}
\label{supp:evaluation}

This supplement provides the evidence needed to interpret and reproduce the experiments in the main paper. It focuses on the common comparison cohort, the semantic-to-geometry interface, additional reconstruction results, and the fixed execution settings. Development logs and superseded pipeline variants are intentionally excluded. Throughout, \(D\) denotes the diagonal of the target axis-aligned bounding box.

\subsection{Targets and Comparison Cohort}

The fixed evaluation cohort contains 20 single-component triangle meshes: sixteen meshes from the Thingi10K dataset cited in the main paper and four pre-existing anonymous reference shapes from our controlled development suite. The Thingi10K cases are \texttt{B1-021}, \texttt{B1-026}, \texttt{B1-048}, \texttt{B3-045}, \texttt{B3-160}, \texttt{B3-212}, \texttt{B3-264}, \texttt{B3-467}, \texttt{B4-0081}, \texttt{B4-0287}, \texttt{B4-0400}, \texttt{B4-0502}, \texttt{B4-0510}, \texttt{B4-0854}, \texttt{B4-0906}, and \texttt{B4-0992}. The four reference shapes are \texttt{C\_7PF2KJ}, \texttt{C\_9RT3WD}, \texttt{DISH}, and \texttt{KNIFE}. The cohort spans smooth shells, thin elongated parts, shallow enclosures, curved ergonomic forms, boundaries, apertures, and cases that stress generic quadrangulation. Source and license metadata are retained with each target; no method receives a source SubD cage or design label.

All five configurations produce comparable outputs on the same 20 targets. Geometry, normal, and silhouette metrics therefore use an identical denominator throughout the main comparison.

\begin{table*}[t]
\centering
\small
\setlength{\tabcolsep}{5pt}
\begin{tabularx}{0.96\textwidth}{@{}lYY@{}}
\toprule
Configuration & Geometry execution & Planning and feedback \\
\midrule
Automatic Remeshing & Complexity-scaled QuadriFlow cage followed by CC-L3 evaluation & None \\
Deterministic Fitting & The same initializer followed by fixed broad and feature refinement & None \\
Rule-based Controller & The same builders, fitters, evidence tools, and gates & Public fixed rules \\
One-shot AI Planning & Planner-selected initialization and feature plan, then fixed execution & One pre-build multimodal decision \\
Ours & The same Stage-A geometry path plus evidence and repair tools & Pre-build planning followed by stateful post-build review \\
\bottomrule
\end{tabularx}
\caption{Compared configurations. Geometry-producing tools are shared wherever applicable; the main difference is how pre-build choices and post-build actions are selected.}
\label{tab:supp-configurations}
\end{table*}

\subsection{Aligned Evidence and Fixed Evaluation}

Every target is observed from six canonical cameras. Each camera provides five spatially aligned evidence modalities: clay appearance, silhouette, world-space normal, dihedral variation, and numbered candidate feature curves. Candidate colors and identifiers remain consistent across views. The planner also receives normalized curve descriptors and topology summaries. Thus, all visual evidence enters the planner at the same stage; candidate-curve images are not treated as a privileged extra channel.

Table~\ref{tab:supp-protocol} summarizes the settings shared by the comparison. One-shot AI Planning and Ours use GPT-5.6 Sol at high reasoning effort in the main comparison. The cross-model study changes only the planner.

\begin{table}[t]
\centering
\small
\setlength{\tabcolsep}{4pt}
\begin{tabularx}{\columnwidth}{@{}lY@{}}
\toprule
Item & Fixed setting \\
\midrule
Input evidence & Six cameras, each with five aligned modalities \\
Initial cage & Generic complexity-scaled QuadriFlow wrapper \\
Requested quad range & 128--384 before the bounded resolution multiplier \\
Surface evaluation & Blender Catmull--Clark at level 3; level 4 used for consistency checks \\
Surface sampling & 20,000 fixed area-weighted samples per surface \\
Distance scale & Target diagonal \(D\) \\
Stage-B budget & At most 15 reviews, 10 geometry attempts, and 600 seconds \\
Checkpoint policy & A proposal replaces the current best only after verification \\
\bottomrule
\end{tabularx}
\caption{Principal evaluation settings. More detailed tool parameters are given in Sec.~\ref{supp:implementation}.}
\label{tab:supp-protocol}
\end{table}

\subsection{Metrics}

For a reconstructed surface \(S\) and target \(T\), directional surface distances are computed from fixed area-weighted samples to the opposite triangle surface. Chamfer-L1 is the mean of the two directional mean distances, normalized by \(D\). Dev.~P95 is the 95th percentile of the reconstruction-to-target distances and is the quantity visualized by the deviation panels. Surface F-score uses precision and recall within \(1\%D\). Normal P95 is the 95th percentile of oriented normal disagreement. Silhouette IoU is averaged over the six canonical views.

A verified checkpoint passes evaluated-surface validity and target topology/aperture preservation. The structural-quality profile additionally reports distance, normal, feature-flow, and pole placement. Feature flow and pole placement are reported as structural diagnostics rather than measurements of user editing performance. A structural warning does not retroactively turn a geometrically valid checkpoint into a topological failure.

\section{Additional Quantitative and Qualitative Results}
\label{supp:additional-results}

\subsection{Additional Method Comparisons}

\begin{figure*}[p]
\centering
\includegraphics[width=\textwidth,height=0.90\textheight,keepaspectratio]{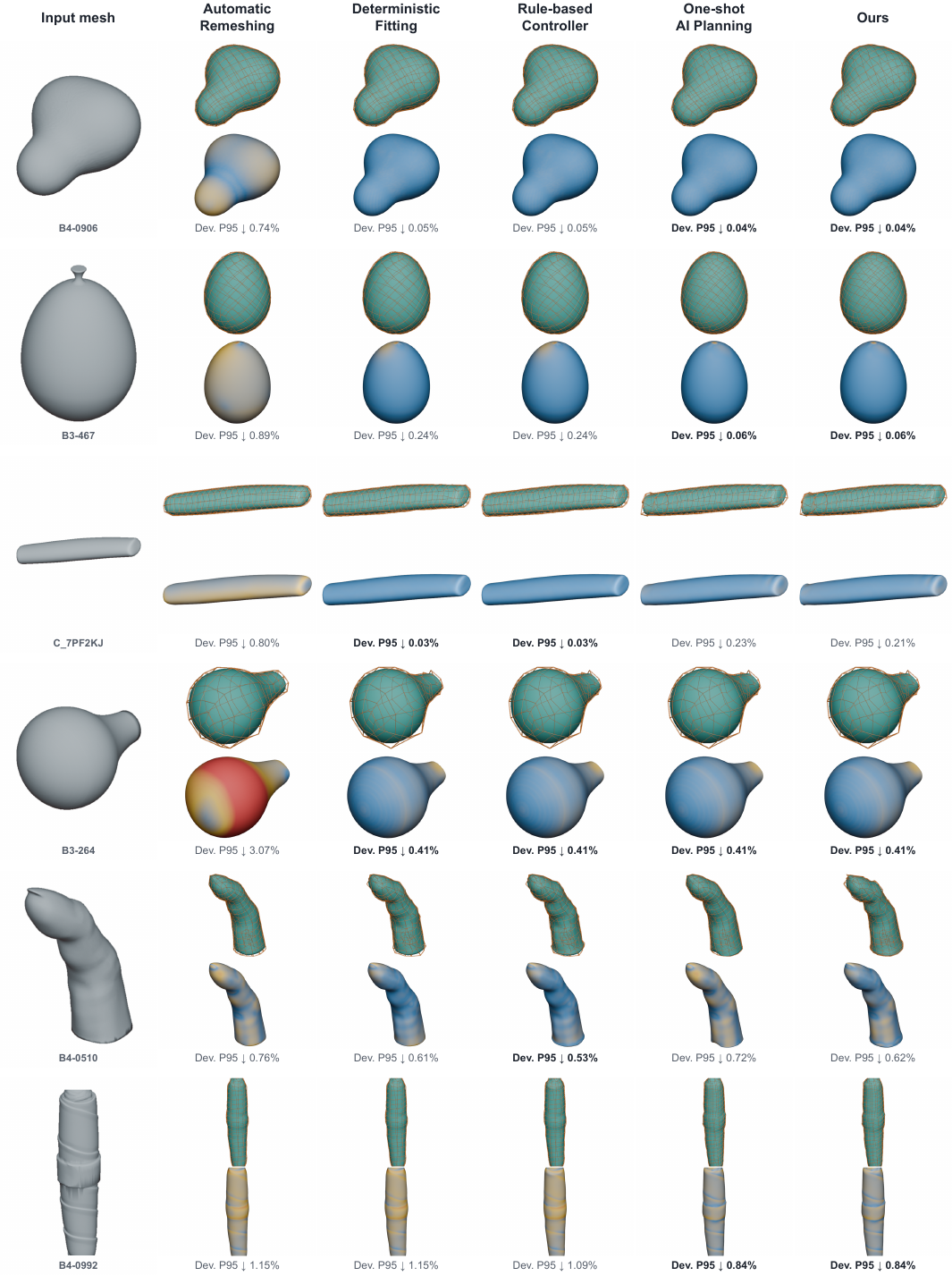}
\caption{Six additional comparisons from the 20-case cohort. Every method cell shows the retained control cage and CC surface together with reconstruction-to-target deviation on a shared \(0\)--\(2\%D\) scale. The displayed Dev.~P95 is the 95th percentile of the same directional distance used in the main table.}
\label{fig:supp-method-extra6}
\end{figure*}

Figure~\ref{fig:supp-method-extra6} extends the main-paper visualization with six additional targets from the same 20-case cohort. Each cell exposes both the control cage and the evaluated CC surface rather than showing a shaded surface alone. The lower panel uses one shared deviation scale, allowing local fitting differences to be compared directly across methods.

\subsection{Cross-Model Geometry}

\begin{figure*}[p]
\centering
\includegraphics[width=\textwidth,height=0.90\textheight,keepaspectratio]{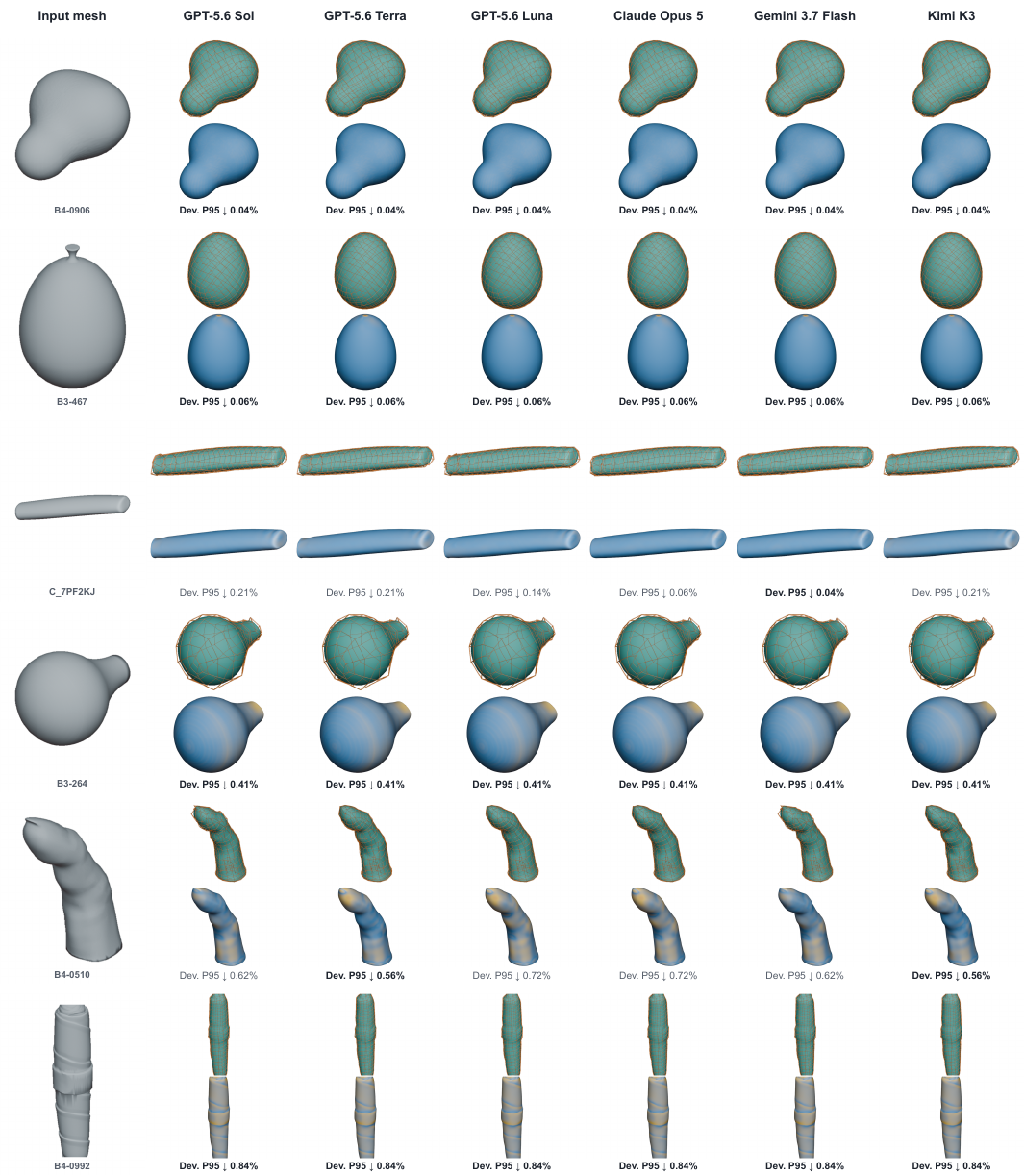}
\caption{Cross-model outputs under the same geometry workspace and verification rules. Repeated rows show cases where planner variation does not change the delivered cage; differing rows show cases where a requested repair or checkpoint choice changes the final geometry.}
\label{fig:supp-crossmodel-extra6}
\end{figure*}

The planner study holds targets, aligned observations, tools, budgets, and deterministic verification fixed. Figure~\ref{fig:supp-crossmodel-extra6} shows six representative targets across GPT-5.6 Sol, GPT-5.6 Terra, GPT-5.6 Luna, Claude Opus 5, Gemini 3.7 Flash, and Kimi K3. Four rows end with identical final cages across all planners; two rows contain several distinct checkpoints. The former illustrate the stabilizing effect of the bounded tool interface, while the latter show where planner choices have a visible geometric consequence.

\clearpage

\section{From Visual Evidence to Subdivision Geometry}
\label{supp:semantic-geometry}

\subsection{A Bounded Modeling Plan}

Stage A converts aligned geometric evidence into a small modeling plan. The planner chooses one control-density tier, one feature-preservation policy, and one semantic action for every numbered candidate curve. Table~\ref{tab:supp-stagea-fields} shows how these decisions enter the geometry workspace.

\begin{table*}[t]
\centering
\small
\setlength{\tabcolsep}{5pt}
\begin{tabularx}{0.96\textwidth}{@{}lYY@{}}
\toprule
Planner decision & Allowed choices & Geometry consequence \\
\midrule
Control density & Low, medium, or high & Scales the complexity-derived requested quad budget before QuadriFlow \\
Feature policy & Smooth, standard, or feature-preserving & Selects a fixed sharp-feature sensitivity preset for initialization \\
Candidate meaning & Hard line, soft ridge, smooth region, boundary, or uncertain & Determines whether the curve can guide fitting, receive a crease, preserve a boundary, or be ignored \\
Candidate action & Real crease, smooth preservation, boundary preservation, or no action & Activates the corresponding mapper or fitter behavior \\
\bottomrule
\end{tabularx}
\caption{Stage-A output and its geometric effect. The planner selects from a fixed set of alternatives; numerical construction and optimization remain tool-controlled.}
\label{tab:supp-stagea-fields}
\end{table*}

The same candidate identifier and color are maintained in all six projections. This matters for occluded or view-dependent features: a curve can be checked against the corresponding clay, silhouette, normal, and dihedral evidence without relying on a single projection. Normalized curve length, endpoints, closure/branch status, and per-view visibility supplement the images. The recorded Mouse example in the main paper shows this planning stage; it is an independent mechanism example and is not part of the 20-case quantitative cohort.

\subsection{Ordered Feature-to-Cage Mapping}

A semantic hard-line decision affects the surface only after it is bound to real cage edges. Let \(P=(p_1,\ldots,p_n)\) be the ordered target feature polyline and let \(G_C\) be the edge graph of the current quad cage. The mapper finds a continuous edge chain whose cost combines three-dimensional curve distance, target-curve progress, tangent agreement, backtracking penalty, and excessive progress-jump penalty. The progress terms prevent a path from taking a geometrically short but semantically wrong route across the rear of the object.

The binding is executed twice:

\begin{enumerate}
    \item On the initial cage, mapped samples become \emph{soft feature anchors}. Broad fitting may move the cage while encouraging its edge flow to follow the target curve.
    \item After broad fitting, the curve is mapped again because the cage has moved. A confirmed hard line is then written as crease weights on the verified cage-edge chain; a soft ridge keeps the positional/tangent objective without hardening the surface.
\end{enumerate}

An unmappable curve is reported and omitted; the mapper never substitutes a disconnected or non-edge path. It neither inserts arbitrary connectivity nor moves control vertices, but selects an ordered path from edges already present in the initial cage.

In this trace, five accepted updates refer to optimization iterations, not five independently detected features. The example is included because it exposes a nonzero, inspectable causal chain:
\[
\begin{aligned}
\text{aligned evidence} &\rightarrow \text{feature semantics} \\
&\rightarrow \text{ordered cage-edge paths} \\
&\rightarrow \text{anchors/creases} \\
&\rightarrow \text{CC-L3 geometry}.
\end{aligned}
\]

\section{Stateful Feedback and Implementation Details}
\label{supp:implementation}

\subsection{Post-Build Actions}

After the initial reconstruction, the Agent reviews the current CC-L3 surface, cage overlay, aligned target/current evidence, recorded metrics, available checkpoints, and the previous tool outcome. It then selects one action from the menu in Table~\ref{tab:supp-actions}. The menu is regenerated from the current state, so an unavailable tool or exhausted budget cannot be requested as an executable action.

\begin{table*}[t]
\centering
\small
\setlength{\tabcolsep}{5pt}
\begin{tabularx}{0.96\textwidth}{@{}lYY@{}}
\toprule
Action family & Agent decision & Tool execution \\
\midrule
Acquire evidence & Request a current multiview comparison, residual close-up, or exact section & Render the requested diagnostic and add it to the next review \\
Regional refit & Select the full cage or a symbolic normalized-height region & Run the fixed fitter and retain the proposal only if verified and improved \\
Support-loop repair & Invoke the loop tool when an executable site exists & Insert one finite-width support loop, locally fit it, and evaluate the new cage \\
Checkpoint navigation & Revisit the broad or seed state, or restore the previous verified state & Load and re-evaluate the checkpoint \\
Finish & Accept the best verified state or abstain & Re-evaluate saved checkpoints and terminate without overriding hard geometry checks \\
\bottomrule
\end{tabularx}
\caption{Stateful Agent action space. The Agent chooses what operation to try; geometry tools construct and verify its consequence.}
\label{tab:supp-actions}
\end{table*}

\subsection{Complementary Controlled Program Switch}

The 20-shape cohort has no source subdivision cages. We therefore retain one compact controlled example with a sealed source surface to show that a planner decision can change the executed construction program. The source is unavailable during planning and is opened only for post-terminal visualization.

The first program produces a geometrically valid but target-inaccurate paired-wall cage. Its evaluation is returned to the planner, which switches to a different construction program. The second cage passes the target-quality criterion. Table~\ref{tab:supp-switch} reports the measured change. This controlled example is reported separately from the 20-case evaluation and is not used to claim broad program-selection superiority.

\begin{table}[t]
\centering
\small
\setlength{\tabcolsep}{4pt}
\begin{tabular}{@{}lrrl@{}}
\toprule
Construction & Mean & P95 & Outcome \\
\midrule
First branch & 3.456\%\(D\) & 7.827\%\(D\) & Rejected \\
Switched branch & 1.348\%\(D\) & 2.952\%\(D\) & Accepted \\
\bottomrule
\end{tabular}
\caption{Controlled rejection--switch--rebuild example. The builders and fitters produce both rows; the planner selects the branch transition. Mean and P95 denote symmetric surface-distance statistics normalized by \(D\).}
\label{tab:supp-switch}
\end{table}

\subsection{Deterministic Construction and Verification}

The generic initial cage uses the complexity-derived budget
\[
Q_{\mathrm{base}}=\operatorname{round}_{32}
\left(128+16\min(C,8)+16\left\lceil\min(L,4)\right\rceil\right),
\]
clamped to \([128,384]\), where \(C\) is the number of candidate curves and \(L\) is their total normalized length. Low, medium, and high planning tiers multiply this budget by \(0.75\), \(1.0\), and \(1.5\). Smooth, standard, and feature-preserving policies use fixed sharp-angle presets of \(50^\circ\), \(35^\circ\), and \(30^\circ\), respectively. If the selected initialization is unusable, the controller may try one predefined lower tier; it does not perform a parameter sweep.

Broad fitting combines robust surface distance, reverse coverage, point-to-plane, normal, silhouette, feature-anchor, displacement, and Laplacian terms. Feature refinement reuses the broad cage with stronger point/tangent terms and the post-fit feature mapping. All final surfaces are reconstructed independently in Blender with the recorded crease and boundary attributes.

Hard rejection is reserved for evaluated-surface self-intersection, fold or orientation/Jacobian failure, degeneracy, and target topology or aperture corruption. Distance, normal, feature flow, and pole placement are reported by the structural-quality profile. These are structural diagnostics, not measurements of user editing performance. A repair is committed only when it passes verification and improves score $S$; otherwise the previous checkpoint remains authoritative.

\subsection{Scope}

The main limitation is the generic initializer: fixed-topology fitting and local repair require a usable starting cage and cannot recover from fundamentally missing components or arbitrary edge flow. The current Agent can request evidence, run regional fitting, invoke one loop operation, and navigate checkpoints, but it cannot invent an unrestricted topology or decompose an object into independently joined parts. The study also does not yet measure professional editing time or compare against commercial retopology workflows. These are directions for broader evaluation, outside the claims of the present system.

\end{document}